\documentclass[a4paper,11pt]{article}
\pdfoutput=1
\usepackage{jheppub,MnSymbol}

\newcommand{\beq}{\begin{equation}}
\newcommand{\eeq}{\end{equation}}
\newcommand{\bea}{\begin{eqnarray}}
\newcommand{\eea}{\end{eqnarray}}

\newcommand{\sle}{\tilde{\ell}}

\preprint{MS-TP-18-03}

\title{Slepton pair production at the LHC in NLO+NLL with resummation-improved parton densities}

\author[a]{\!\!Juri Fiaschi}
\author[a]{\!\!and Michael Klasen}

\affiliation[a]{Institut f\"ur Theoretische Physik, Westf\"alische Wilhelms-Universit\"at
 M\"unster, Wilhelm-Klemm-Stra\ss{}e 9, D-48149 M\"unster, Germany}

\emailAdd{fiaschi@uni-muenster.de}
\emailAdd{michael.klasen@uni-muenster.de}

\abstract{Novel PDFs taking into account resummation-improved
matrix elements, albeit only in the fit of a reduced data set,
allow for consistent NLO+NLL calculations of slepton pair
production at the LHC. We apply a factorisation method to this
process that minimises the effect of the data set reduction,
avoids the problem of outlier replicas in the NNPDF method for
PDF uncertainties and preserves the reduction of the scale
uncertainty. For Run II of the LHC, left-handed selectron/smuon,
right-handed and maximally mixed stau production, we confirm
that the consistent use of threshold-improved PDFs partially
compensates the resummation contributions in the matrix elements.
Together with the reduction of the scale uncertainty at NLO+NLL,
the described method further increases the reliability of slepton
pair production cross sections at the LHC.
}


\begin{document}
\maketitle
\flushbottom

\section{Introduction}

Supersymmetry (SUSY) is a well-motivated extension of the Standard
Model (SM) of particle physics that allows to answer simultaneously a
significant number of its open questions \cite{Nilles:1983ge,%
Haber:1984rc} , but that can also be simplified in ways that can serve
as prototypes for other, differently motivated models that address only
a subset of these questions \cite{Alwall:2008ag,Calibbi:2014lga}.
Consequently, the search for SUSY particles continues to be an
important research objective at the CERN LHC \cite{Giordano:2017dqe}.
For the determination or exclusion of SUSY cross sections, particle
masses and parameters \cite{AguilarSaavedra:2005pw}, the experiments
ATLAS and CMS rely on precise theoretical predictions in next-to-leading
order (NLO) QCD \cite{Beenakker:1996ch,Beenakker:1997ut,Beenakker:1999xh,%
Berger:1999mc,Berger:2000iu,Spira:2002rd,Jin:2003ez,Binoth:2011xi} and beyond.
In particular, threshold resummation techniques have been used
successfully to take into account the emission of soft gluons and the
associated leading, next-to-leading (NLL) and even
next-to-next-to-leading logarithmic corrections to all orders in the
strong coupling constant $\alpha_s$. These calculations have been performed for
the production of squarks and gluinos \cite{Beenakker:2014sma,%
Borschensky:2014cia,Beneke:2016kvz,Beenakker:2016lwe}, stops
\cite{Broggio:2013cia,Beenakker:2016gmf}, gauginos \cite{Li:2007ih,%
Debove:2010kf,Debove:2011xj,Fuks:2012qx,Fuks:2013vua,Fuks:2016vdc} and
sleptons \cite{Yang:2005ts,Broggio:2011bd,Bozzi:2007qr,Bozzi:2007tea,%
Fuks:2013lya}.

In our previous work on slepton pair production processes, we computed
their NLO SUSY-QCD corrections \cite{Beenakker:1999xh} as well as NLL
transverse-momentum \cite{Bozzi:2006fw}, threshold \cite{Bozzi:2007qr}
and joint \cite{Bozzi:2007tea} resummation corrections. These results
are regularly applied to experimental analyses at the LHC
\cite{Aad:2015eda,Aaboud:2017leg,Khachatryan:2014qwa,CMS:2017wox,CMS:2017mkt}.
We subsequently released the public code RESUMMINO \cite{Fuks:2013vua},
applied our results to simplified models and reanalysed public ATLAS
and CMS data for various assumptions on the decomposition of the
sleptons and their neutralino decay products \cite{Fuks:2013lya}.

The NLO+NLL corrections often increase the total SUSY particle
production cross sections and therefore the experimental sensitivity
with respect to NLO calculations. More importantly, they allow for a
reduction of the theoretical uncertainty from the variation of the
unphysical renormalisation and factorisation scales. While this could
also be demonstrated in our previous work for slepton pair production,
the second important theoretical uncertainty coming from the spread of
parton density functions (PDFs) was not reduced, as they were only
determined by fitting data from deep-inelastic scattering (DIS), the
Drell-Yan (DY) process, jet production etc.\ using NLO partonic matrix
elements \cite{Lai:2010vv,Martin:2009iq,Ball:2013hta}.

For consistency, hadronic cross sections should be computed with the
same level of accuracy in both the partonic cross section and the PDFs.
With the publication of the NNPDF30\_nll\_disdytop PDFs \cite{Bonvini:2015ira},
this has now become possible at the NLO+NLL level. However, since only
a reduced number of partonic cross sections that usually enter global
fits are available with NLO+NLL precision, only a subset of the
corresponding data sets, i.e.\ on DIS, DY and top pair production,
could be used in the determination of these threshold-resummation improved PDFs. The
consequence is that the PDF uncertainty band, produced with the NNPDF
replica method, is actually larger, not smaller than with globally
fitted NLO PDFs. One of the examples used in Ref.\ \cite{Bonvini:2015ira}
is in fact slepton pair production, calculated with RESUMMINO. However,
only one invariant-mass distribution for left-handed selectrons of mass
564 GeV is studied there.

The NNPDF30\_nll\_disdytop PDFs have subsequently been employed to investigate
the effect of threshold-resummation improved PDFs on squark and gluino
production \cite{Beenakker:2015rna}. The authors showed that the total
central cross sections were modified both in a qualitative and quantitative
way, illustrating the relevance and impact of using threshold-resummation
improved PDFs. They introduced in particular a factorisation
($K$-factor) method, described in detail below, that allows to combine
the impact of NLO+NLL threshold resummation on the reduced central PDF
fit with the smaller uncertainty of the global NLO PDF fit, leading to
approximately consistent NLO+NLL hadronic cross sections. This method
also allows to avoid the problem of exceedingly large, small or even
negative cross sections induced by outlier replicas and the lack of
positivity constraints on the NNPDF PDFs. The replica method is known
to be particularly problematic for resummation calculations, which rely
on a transformation of PDFs to Mellin space from all regions of $x$,
including those, where they are not well constrained \cite{Fuks:2013lya}.

The purpose of this paper is therefore threefold. First, our NLO+NLL
predictions for slepton pair production are updated to the current
LHC collision energy of 13 TeV. Second, we employ with NNPDF3.0 an
up-to-date global set of PDFs, that is now also based on ATLAS and
CMS data from jet, vector-boson and top-quark production
\cite{Ball:2014uwa}. This allows us to reach our third and central
goal of studying the impact of threshold-resummation improved PDFs
not only on differential, but also on total slepton pair production
cross sections and for a variety of SUSY scenarios. 

The remainder of this paper is organised as follows: In Sec.\
\ref{sec:2} we describe our theoretical approach using the $K$-factor
method and how we combine NLO+NLL resummation effects with global
PDF and also scale uncertainties. In Sec.\ \ref{sec:3}, we present
numerical results for differential and total cross sections of
left-handed first and second generation sleptons in graphical and
tabular form. We do this for $pp$ collisions of 13 TeV
centre-of-mass energy and various slepton masses relevant for Run
II of the LHC. Similar results are presented in Sec.\ \ref{sec:4}
for third-generation sleptons, i.e.\ right-handed or maximally
mixed staus. Our conclusions are given in Sec.\ \ref{sec:5}.

\newpage
\section{Theoretical method}
\label{sec:2}

Apart from updating our NLO+NLL predictions for slepton pair production
to the current experimental conditions at Run II of the LHC, i.e.\
to proton-proton collisions with 13 TeV centre-of-mass energy, and with
recent PDF sets from the global NNPDF3.0 fits, which are now also based
on ATLAS and CMS data from jet, vector-boson and top-quark production
\cite{Ball:2014uwa}, the central goal of this work is to quantify the
impact of threshold-resummation improved PDFs on our predictions.
These PDFs, called NNPDF30\_nll\_disdytop, have only recently been made available
by the NNPDF collaboration \cite{Bonvini:2015ira}. They are based on a
similar setup as those at leading order (LO) and NLO, but use partonic matrix elements
at NLO+NLL, albeit so far only for a smaller set of processes
(Deep-Inelastic Scattering, Drell-Yan and top-quark pair production),
for which these matrix elements at NLO+NLL are available. Together with
this NNPDF30\_nll\_disdytop fit, an NLO fit (NNPDF30\_nlo\_disdytop)
based on the same subsample of processes and data sets has been
provided. Unfortunately, the reduction of the input data set reduces
the precision of these fits, and consequently they have currently still
larger uncertainties than the global sets.

Threshold-resummation improved PDFs have previously been applied to
squark and gluino production cross sections at NLO+NLL. The result
there was that their effect cannot be neglected, as it modifies both the
qualitative and the quantitative behaviour of the sparticle pair
production cross sections \cite{Beenakker:2015rna}. In order to
eliminate the impact of the reduction of the fitted data set, the authors
introduced a $K$-factor
\begin{equation}
 K =
 \frac{\sigma({\rm NLO+NLL})_{\rm NLO~global}}{\sigma({\rm NLO})_{\rm NLO~global}}
 \cdot
 \frac{\sigma({\rm NLO+NLL})_{\rm NLO+NLL~reduced}}{\sigma({\rm NLO+NLL})_{\rm NLO~reduced}},
 \label{eq:K_factor}
\end{equation}
which allows to obtain (approximate) central total NLO+NLL cross
sections with NLO+NLL PDFs via
\begin{equation}
 \sigma({\rm NLO+NLL})_{\rm NLL+NLO~global} = K \cdot \sigma({\rm NLO})_{\rm NLO~global}.
 \label{eq:K_application}
\end{equation}
Varying the NLO global PDFs in $\sigma({\rm NLO})_{\rm NLO~global}$ with
their reliable spread then produces a reliable (approximate) NLO+NLL
global PDF error. In Eq.\ (\ref{eq:K_factor}), the first ratio takes
into account the effect of the resummation in the partonic matrix
elements using the NLO PDF fit of the global data set, whereas the
second ratio parameterises the impact of the threshold-resummation
improved partonic matrix elements on the NLO+NLL PDF fit of the reduced
data set. As explained above, we use the NLO global NNPDF3.0 PDF set
\cite{Ball:2014uwa} for the computation of the first ratio related to
the global part, while the NLO+NLL and NLO PDF fits based on the
reduced data set \cite{Bonvini:2015ira} enter the evaluation of the
second ratio. A definition equivalent to Eq.\ (\ref{eq:K_application})
is adopted for invariant-mass distributions by simply replacing the
total integrated cross section $\sigma$ with the differential cross
sections $d\sigma/dM_{\sle\sle}$.

With the method described above, we also bypass a known issue with the
NNPDF approach to PDF uncertainties. In order to compute this
systematic error, the NNPDF collaboration fit a large number of
Gaussian-distributed replicas of the experimental data without imposing
ad-hoc conditions on the shape or positivity of the PDFs to avoid
theoretical bias or underestimation of the resulting PDF uncertainty.
The PDF uncertainty on any observable results from applying to it the
ensemble of the replica fits. However, positivity is and practically
can be checked only for a subset of observables. The complication
encountered in resummation calculations is that the PDF replicas have
to be transformed to Mellin space, i.e.\ be integrated over regions
in $x$ where they are not well constrained, which can then lead to
unphysically large variations of the resummed cross sections
\cite{Fuks:2013lya}. At NLO and in $x$-space, one can directly
eliminate the replicas that feature such misbehaviour
\cite{Ball:2014uwa}. Alternatively, one can consider only the 68\%
CL interval by eliminating the replicas leading to the lowest and
highest 16 cross sections, respectively, and use the midpoint of this
interval as the central prediction \cite{Butterworth:2015oua}.
Note that this prescription leads to central predictions
that differ substantially from those obtained with the central PDF fit.
The $K$-factor method described above provides a more elegant
solution to the problem of outlier replicas entering resummation
predictions by avoiding their transformation to Mellin space altogether.
According to Eq.\ (\ref{eq:K_factor}) only the NLO global, NLO reduced
and NLO+NLL reduced central fits have to be transformed, and according
to Eq.\ (\ref{eq:K_application}) the global replicas have to be applied
only at NLO and in $x$-space.

The $K$-factor method conceals one important benefit of the resummation
calculation, which is the sizeable reduction of scale uncertainties.
This reduction from NLO to NLO+NLL would be lost if the scales were
varied only in the NLO cross section in Eq.\ (\ref{eq:K_application})
with respect to the central scale $\mu_R=\mu_F=m_{\sle}$. It is therefore
evaluated directly in the NLO+NLL (or NLO) cross sections by applying
the usual seven-point method of relative factors of two, but not four
among the two types of scales. The total theoretical uncertainty is then
obtained by adding the relative PDF and scale uncertainties in quadrature.

\section{Left-handed selectron/smuon pair production}
\label{sec:3}

In this section, we study the effects of the threshold-improved NLO+NLL
PDFs as implemented in LHAPDF6 \cite{Buckley:2014ana} on invariant-mass
distributions and total cross sections for left-handed first- and
second-generation slepton pair production using RESUMMINO
\cite{Fuks:2013vua}. If we assume as usual universality of the
corresponding soft SUSY-breaking masses and do not take into account
branching ratios and experimental efficiencies, selectron and smuon
production cross sections are identical. We set all SM parameters to
their current PDG values \cite{Patrignani:2016xqp} and use
$\alpha_s(M_Z)=0.118$ with $\Lambda_{n_f=5}^{\overline{\rm MS}}=0.239$ GeV as
appropriate for NNPDF3.0.

\begin{figure}
\begin{center}
\includegraphics[width=\textwidth]{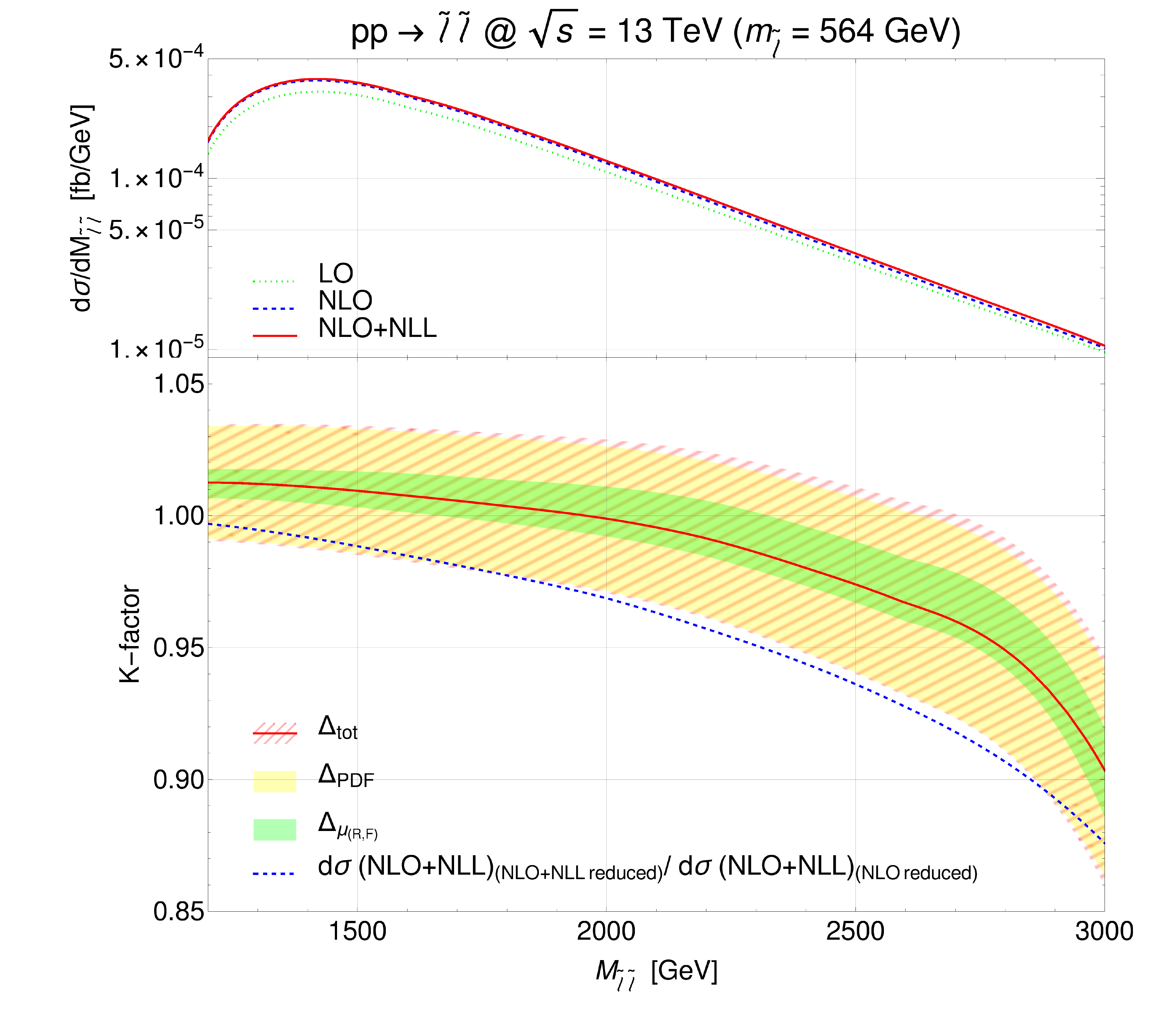}
\caption{Invariant-mass distributions (upper panel) and $K$-factors
 (lower panel) according to Eq.\ (\ref{eq:K_factor}) using the full
 expression (full red) and only its second, PDF-dependent part (dashed
 blue line) for the pair production of left-handed selectrons/smuons
 with a mass of $m_{\tilde{\ell}} = 564$ GeV at the LHC with $\sqrt{s} = 13$ TeV.
 In the upper panel, the results at LO (dotted green), NLO (dashed blue)
 and NLO+NLL (full red line) have been obtained with global NLO PDFs.
 In the lower panel, the PDF (yellow) and scale (green) uncertainties
 have been computed at NLO and NLO+NLL, respectively, with global NLO
 PDFs, then rescaled appropriately and added in quadrature for the
 total theoretical uncertainty (dashed red).}
\label{fig:inv_mass_slepton}
\end{center}
\end{figure}

In the upper panel of Fig.\ \ref{fig:inv_mass_slepton}, we show the
invariant-mass distribution of left-handed selectron/smuon pairs with
a fixed mass of $m_{\tilde{\ell}} = 564$ GeV. This mass is identical to
the one chosen in Ref.\ \cite{Bonvini:2015ira} in order to facilitate
a straightforward comparison of our results. The invariant mass
distributions, computed at LO (dotted green), NLO (dashed blue) and
NLO+NLL (full red line) in the matrix elements, but always with
global NLO NNPDF3.0 PDFs, exhibit the typical rise above pair
production threshold to about $M_{\sle\sle}=1.4$ TeV and a subsequent
fall-off. At the maximum, an increase of about 16\% is visible from
LO to NLO with an increase of another 2\% from NLO to NLO+NLL, which
then rises to 3\% at $M_{\sle\sle}=3$ TeV as expected.

The lower panel of Fig.\ \ref{fig:inv_mass_slepton} shows the
$K$-factor as defined in Eq.\ (\ref{eq:K_factor}) (full red) as well
as its second part (dashed blue line) that comes from the change of
PDFs alone. The latter amounts to a decrease of more than 10\% at high
invariant mass, which is partially compensated by the NLL corrections
in the matrix elements. At low invariant mass, one observes even an
overcompensation, such that the total $K$-factor is slightly larger
than unity. This effect was also observed in Fig.\ 17 of Ref.\
\cite{Bonvini:2015ira}. Our results agree quite well with theirs
despite the fact that they show the slightly different factor
\begin{equation}
 K' =
 \frac{\sigma({\rm NLO+NLL})_{\rm NLO+NLL~reduced}}{\sigma({\rm NLO})_{\rm NLO~global}}.
\end{equation}
This $K'$-factor thus also includes the impact of the reduction of the
data set in the PDF fits from NLO to NLO+NLL, which we preferred to
remove from our analysis and which impacts mostly the region of large
invariant mass or parton momentum fraction, where the PDFs are not well
constrained. A second difference in our figure is our (yellow) PDF
uncertainty band, which is based on the more reliable global NLO fit,
while the (dashed red) band in Ref.\ \cite{Bonvini:2015ira} is based
on the reduced data set and thus considerably larger. Furthermore, we
also show the (green) scale uncertainty obtained with the usual
seven-point method, which, as expected from the reduction due to
NLL resummation contributions, contributes only little to the total
(dashed red) uncertainty (added in quadrature). Remember that the
scale uncertainty has been computed directly at NLO+NLL using the
global NLO PDFs and has then been rescaled appropriately (cf.\ Sec.\
\ref{sec:2}).

To estimate the size of (N)NNLL over NLL threshold resummation effects
in both the matrix elements and the PDFs, it is instructive to compare
the invariant mass distributions for slepton pairs in the lower right
plot of Fig.\ 7 of Ref.\ \cite{Broggio:2011bd} and of the quark-antiquark
luminosities in the upper left plots of Figs.\ 13 and 14 of Ref.\
\cite{Bonvini:2015ira}. While the former lie at the upper end of the
NLL uncertainty band, the latter are reduced by 1-2\% with respect to
NLL, confirming again further increased perturbative stability and
additional partial compensation of resummation effects in the PDFs.

\begin{figure}
\begin{center}
\includegraphics[width=\textwidth]{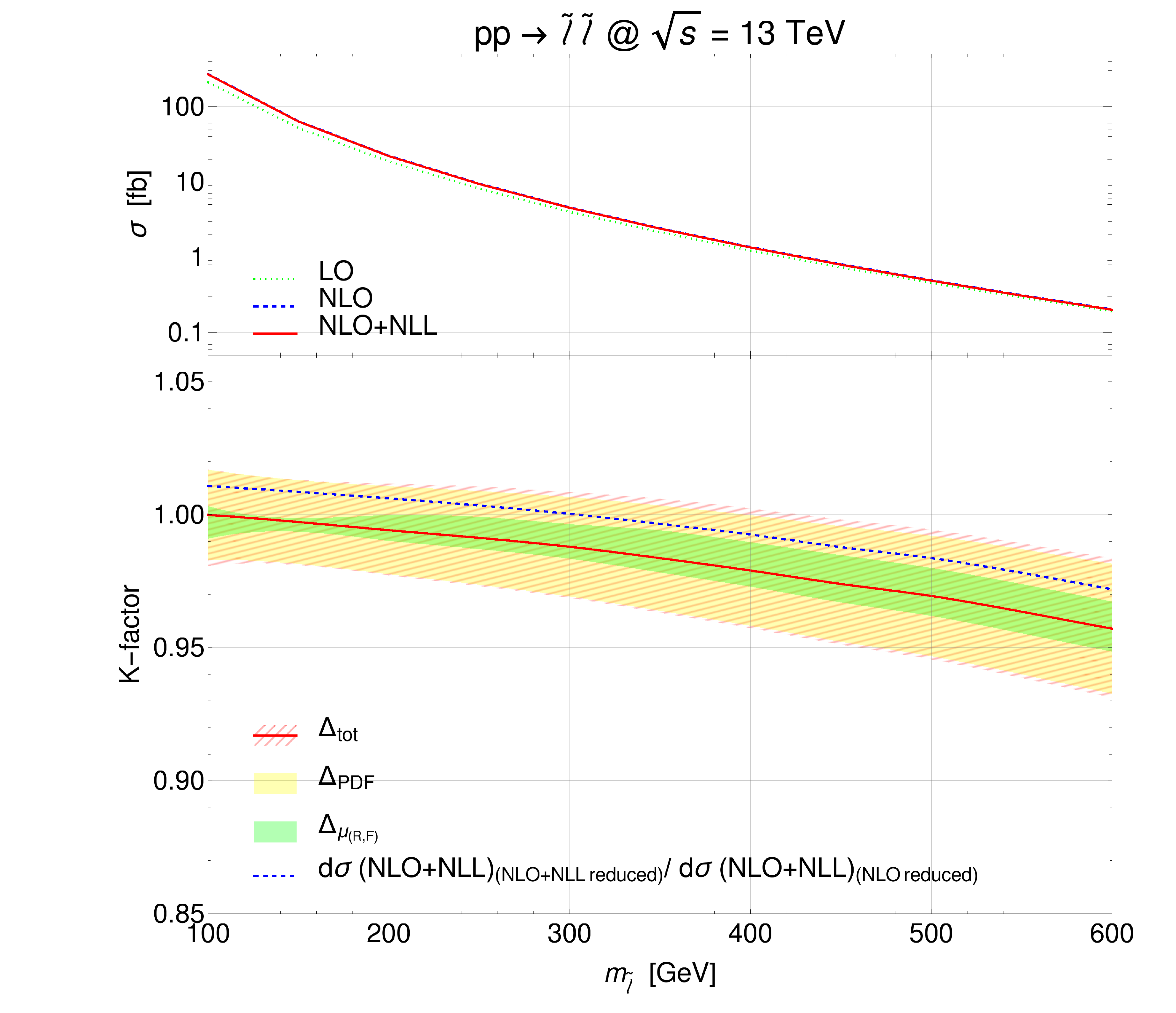}
\caption{Same as Fig.\ \ref{fig:inv_mass_slepton}, but for the
 total cross section as a function of the selectron/smuon mass.}
\label{fig:total_cross_section_slepton}
\end{center}
\end{figure}

In Fig.\ \ref{fig:total_cross_section_slepton} we show similar
results, but now for the total cross section as a function of the
selectron/smuon mass. As can be seen in the upper panel, left-handed
sleptons should have been produced in significant numbers already at
Run II of the LHC with luminosities recorded in 2016-2017 by ATLAS
and CMS of 35-50 fb$^{-1}$ each over most of the mass region shown.
Indeed, current left- (right-) handed slepton mass limits reach
values of 400 (290) GeV,
but they depend strongly on the mass splitting with the lightest
SUSY particle, usually assumed to be the lightest neutralino
$\tilde{\chi}_1^0$ \cite{Aaboud:2017leg,CMS:2017mkt}.

The central $K$-factor (full red line) in the lower panel shows that
the NLO+NLL PDFs reduce not only the invariant-mass distribution, but
also the total cross section by up to 4\% for large slepton masses,
where their effect partially compensates again the impact of the NLL
corrections in the matrix elements. This result agrees with the one
for squark pair production in Fig.\ 8 and with the result for the
underlying quark-quark luminosity in Fig.\ 7 of Ref.\ \cite{Beenakker:2015rna}.
The difference with the quark-antiquark luminosity is of minor importance
in the sea-quark region. For large slepton masses, the total $K$-factor
from NLO to NLO+NLL is not only larger than the scale, but also the PDF
uncertainty. In contrast, the impact of the NLO+NLL matrix elements on
the PDF fit alone falls within this uncertainty.

We conclude this section for future use, e.g.\ by the LHC experiments,
with explicit results in Tab.\ \ref{tab:Selectron_table} on the total
cross sections
\begin{table}
\caption{Total cross section for first-generation slepton pair
 production at the LHC with $\sqrt{s} = 13$ TeV as a function of
 the slepton mass at LO, NLO and NLO+NLL with consistent PDF 
 choices. The central NLO+NLL results are obtained with the
 $K$-factor method, whereas the NLO+NLL (asymmetric) scale
 uncertainty has been computed directly, and the PDF (symmetric)
 uncertainty at NLO (identical in the last two columns).}
\label{tab:Selectron_table}
\begin{center}
\begin{tabular}{|c||c|c|c|}
  \hline
  $m_{\tilde{\ell}}$ [GeV] & LO (LO global) [fb] & NLO (NLO global) [fb] & NLO+NLL (id.\ global) [fb]\\
  \hline
  $100$ & $210.40^{+3.5\%}_{-4.4\%}\pm7.6\%$ & $271.24^{+2.1\%}_{-1.7\%}\pm1.7\%$ & $271.22^{+0.3\%}_{-0.9\%}\pm1.7\%$ \\
  \hline
  $150$ & $52.16^{+0.6\%}_{-1.2\%}\pm6.3\%$ & $64.23^{+2.1\%}_{-1.4\%}\pm1.6\%$ & $64.05^{+0.2\%}_{-0.4\%}\pm1.6\%$ \\
  \hline
  $200$ & $18.68^{+0.8\%}_{-1.1\%}\pm6.2\%$ & $22.32^{+2.1\%}_{-1.5\%}\pm1.7\%$ & $22.19^{+0.6\%}_{-0.4\%}\pm1.7\%$ \\
  \hline
  $250$ & $8.14^{+2.2\%}_{-2.4\%}\pm6.2\%$ & $9.51^{+2.2\%}_{-1.6\%}\pm1.8\%$ & $9.42^{+0.8\%}_{-0.4\%}\pm1.8\%$ \\
  \hline
  $300$ & $4.01^{+3.4\%}_{-3.3\%}\pm6.2\%$ & $4.60^{+2.1\%}_{-1.8\%}\pm1.9\%$ & $4.55^{+0.9\%}_{-0.5\%}\pm1.9\%$ \\
  \hline
  $350$ & $2.15^{+4.3\%}_{-4.1\%}\pm6.2\%$ & $2.43^{+2.1\%}_{-2.0\%}\pm2.0\%$ & $2.39^{+1.1\%}_{-0.5\%}\pm2.0\%$ \\
  \hline
  $400$ & $1.23^{+5.1\%}_{-4.8\%}\pm6.2\%$ & $1.37^{+2.1\%}_{-2.1\%}\pm2.1\%$ & $1.34^{+1.1\%}_{-0.6\%}\pm2.1\%$ \\
  \hline
  $450$ & $0.74^{+5.8\%}_{-5.3\%}\pm6.3\%$ & $0.81^{+2.1\%}_{-2.2\%}\pm2.2\%$ & $0.79^{+1.1\%}_{-0.7\%}\pm2.2\%$ \\
  \hline
  $500$ & $0.46^{+6.5\%}_{-5.8\%}\pm6.4\%$ & $0.50^{+2.1\%}_{-2.3\%}\pm2.3\%$ & $0.48^{+1.1\%}_{-0.8\%}\pm2.3\%$ \\
  \hline
  $550$ & $0.29^{+7.0\%}_{-6.3\%}\pm6.5\%$ & $0.31^{+2.1\%}_{-2.4\%}\pm2.4\%$ & $0.30^{+1.1\%}_{-0.8\%}\pm2.4\%$ \\
  \hline
  $600$ & $0.19^{+7.5\%}_{-6.7\%}\pm6.7\%$ & $0.20^{+2.2\%}_{-2.6\%}\pm2.6\%$ & $0.20^{+1.1\%}_{-0.9\%}\pm2.6\%$ \\
  \hline
\end{tabular}
\end{center}
\end{table}
at LO, NLO and NLO+NLL, that have been obtained with consistent PDF choices
using Eq.\ (\ref{eq:K_application}), and on the
corresponding theoretical uncertainties. The central
NLO+NLL results have been obtained with the $K$-factor method,
while the NLO+NLL (asymmetric) scale uncertainties have been
computed directly, and the PDF (symmetric) uncertainties at NLO.
The latter are therefore identical in the last two columns.

\section{Right-handed and mixed stau pair production}
\label{sec:4}

In this section we repeat the analysis of Sec.\ \ref{sec:3} for
right-handed and maximally mixed stau pair production. Since the
off-diagonal elements of the sfermion mixing matrices are proportional
to the SM fermion mass, mixing of chirality superpartners is only
important for third-generation sfermions, i.e.\ in our case for tau
sleptons. At the cross section level, left-handed stau cross sections
are identical to those for selectrons and smuons, so that we do not
show the corresponding results again. Experimentally, the analysis
for staus is, of course, very different, since their decay products are
unstable tau leptons, that are not directly measured in the tracking
systems, electromagnetic calorimeters or muon chambers, but that have to be
reconstructed themselves from hadronic \cite{Aad:2014yka,CMS-PAS-SUS-17-003}
and/or leptonic \cite{CMS:2017wox} decay products.

\begin{figure}
\begin{center}
\includegraphics[width=.48\textwidth]{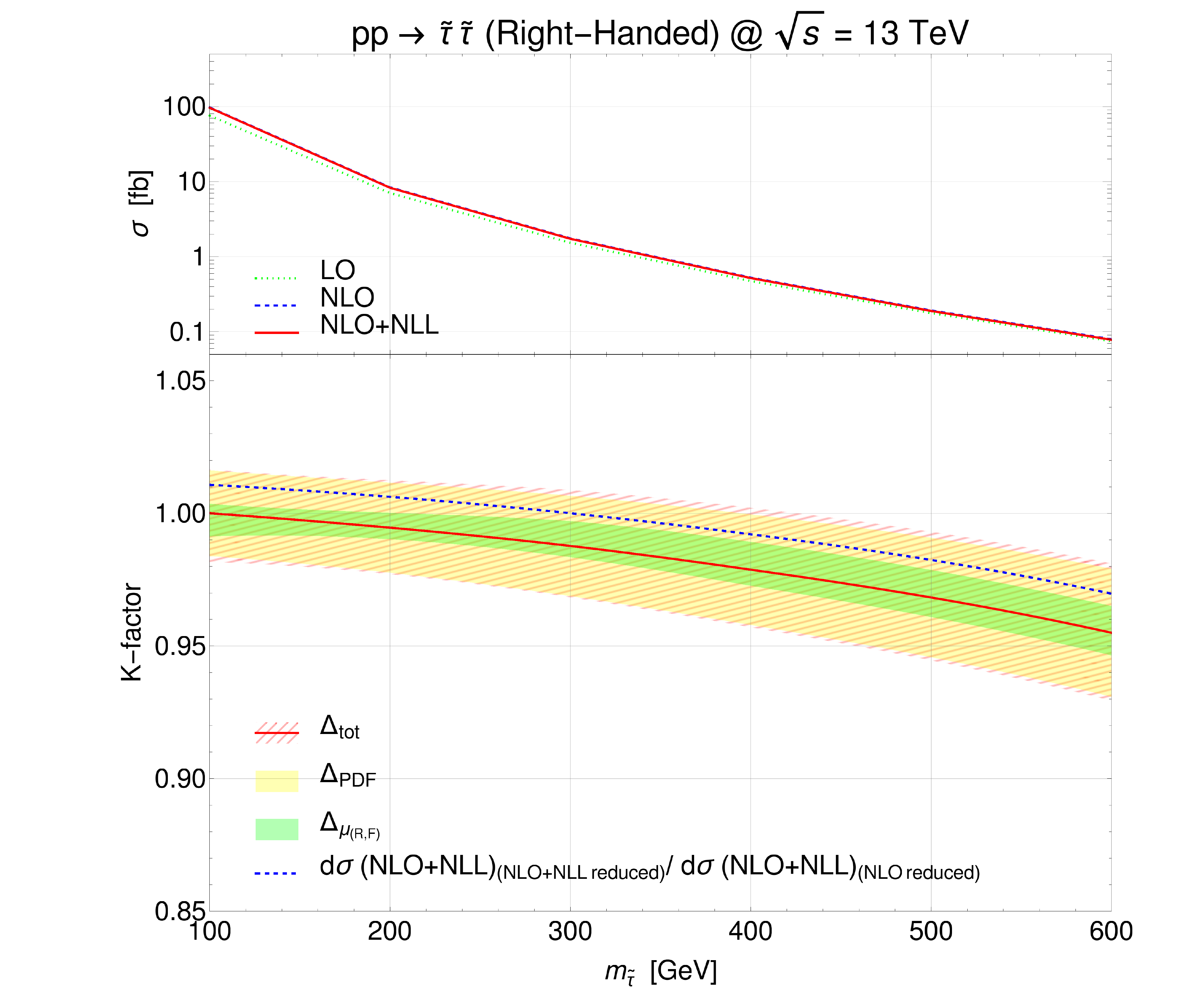}
\includegraphics[width=.48\textwidth]{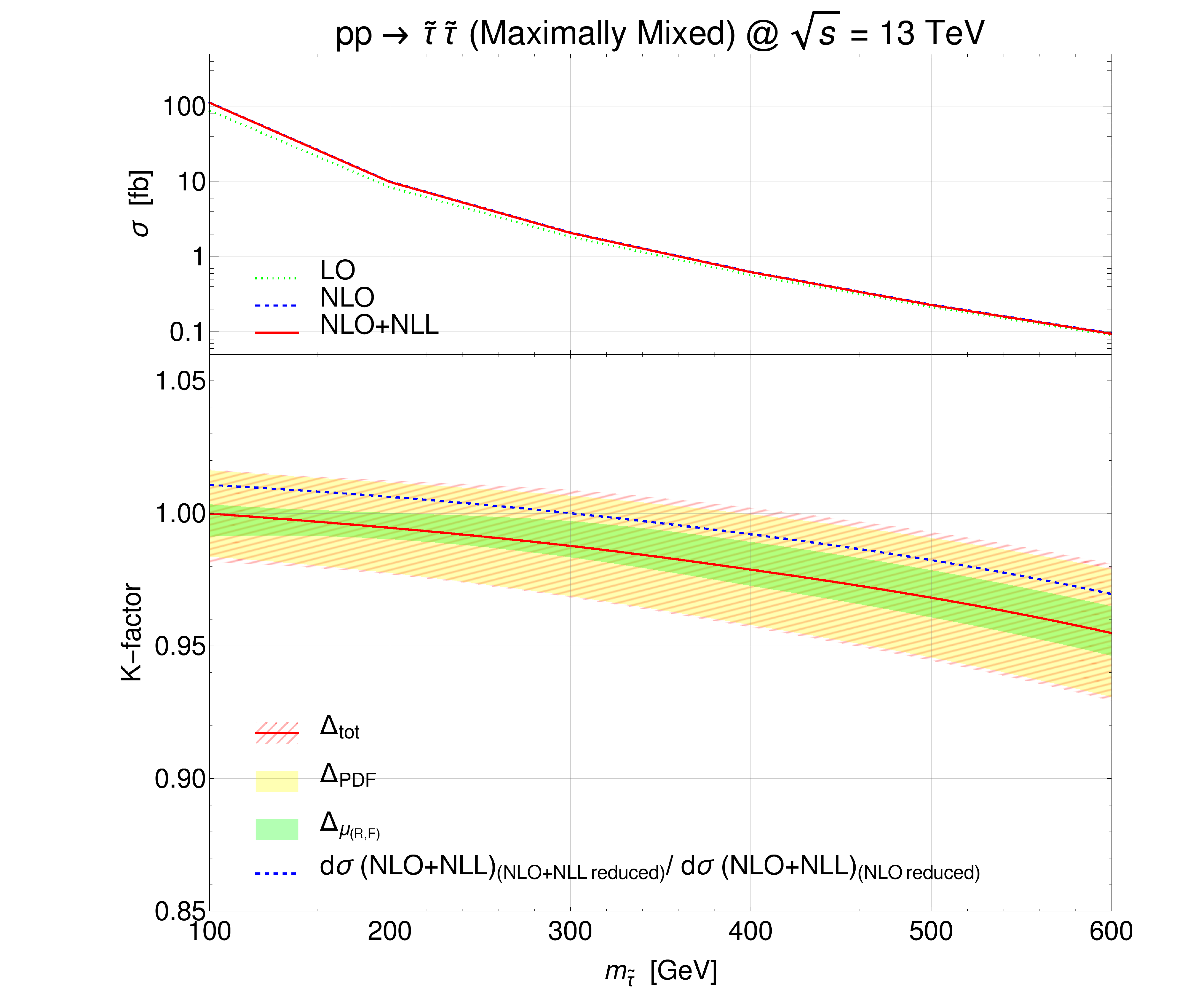}
\caption{Same as Fig.\ \ref{fig:total_cross_section_slepton}, but for the
 pair production of right-handed (left) and maximally mixed (right) staus.}
\label{fig:total_cross_section_stau}
\end{center}
\end{figure}

In the upper panels of Fig.\ \ref{fig:total_cross_section_stau} we
show the total cross sections for right-handed (left) and maximally
mixed (right) stau pairs computed at LO (dotted green), NLO (dashed
blue) and NLO+NLL (full red line) with global NLO PDFs. We follow
here the experimental analysis in Ref.\ \cite{CMS-PAS-SUS-17-003}
and extend it from masses of 400 GeV to 600 GeV. The total cross
sections for right-handed staus are clearly smaller than those for
left-handed sleptons in Fig.\ \ref{fig:total_cross_section_slepton},
with those for maximally mixed staus lying between the two extremes.
Compared to the cross sections in Run I of the LHC at $\sqrt{s}=7$ TeV
(8 TeV) as listed in Tab.\ 1 (2) of Ref.\ \cite{Fuks:2013lya}, they are
increased by a factor of 2.5 (2) at low slepton masses and up to
a factor of 10 (5) at high slepton masses. The PDF update from
the NLO fit of CT10 \cite{Lai:2010vv} used in Ref.\ \cite{Fuks:2013lya}
to the global NLO fit of NNPDF3.0 used here changes the NLO+NLL cross
sections insignificantly at low slepton masses and by up to 5\% at
high slepton masses, which
fell well into the CT10 PDF uncertainty, but exceeds the current
NNPDF3.0 PDF uncertainty. Although the total cross sections are
relatively
large over the full mass range shown and should have led to the
production of staus in significant numbers at the LHC, only upper
limits on the cross sections could so far be derived by ATLAS in
Run I \cite{Aad:2014yka} and by CMS in Run II \cite{CMS-PAS-SUS-17-003}
in the purely hadronic decay channel and by CMS, for left-handed
staus, in the (semi-)leptonic decay channel(s) \cite{CMS:2017wox}.

The lower panels of Fig.\ \ref{fig:total_cross_section_stau}
show the corresponding $K$-factors according to the full expression
of Eq.\ (\ref{eq:K_factor}) (full red) and to its second, PDF-dependent
part only (dashed blue line), together with the PDF (yellow),
scale (green) and total theoretical uncertainty (dashed red).
The QCD corrections turn out to be largely independent of the weak
coupling structure of the underlying partonic cross section, so that
the dependence on the weak couplings cancels in the ratios and
no differences are visible in the $K$-factors with respect to those
of first-generation left-handed sleptons in the lower panel of Fig.\
\ref{fig:total_cross_section_stau}.

For better readibility and future use, we end this section by listing
consistent total cross sections for right-handed and maximally mixed
stau pair production at LO, NLO and NLO+NLL in Tabs.\
\ref{tab:Stau_table_RH} and \ref{tab:Stau_table_MM}.
\begin{table}
\caption{Same as Tab.\ \ref{tab:Selectron_table}, but for the pair
 production of right-handed staus.}
\label{tab:Stau_table_RH}
\begin{center}
\begin{tabular}{|c||c|c|c|}
  \hline
  $m_{\tilde{\tau}}$ [GeV] & LO (LO global) [fb] & NLO (NLO global) [fb] & NLO+NLL (id.\ global) [fb]\\
  \hline
  $100$ & $75.68^{+3.4\%}_{-4.3\%}\pm6.4\%$ & $97.19^{+2.1\%}_{-1.7\%}\pm1.6\%$ & $97.20^{+0.4\%}_{-0.9\%}\pm1.6\%$ \\
  \hline
  $200$ & $7.03^{+0.8\%}_{-1.1\%}\pm6.2\%$ & $8.37^{+2.2\%}_{-1.5\%}\pm1.7\%$ & $8.33^{+0.6\%}_{-0.4\%}\pm1.7\%$ \\
  \hline
  $300$ & $1.53^{+3.3\%}_{-3.3\%}\pm6.2\%$ & $1.75^{+2.1\%}_{-1.8\%}\pm1.9\%$ & $1.73^{+0.9\%}_{-0.4\%}\pm1.9\%$ \\
  \hline
  $400$ & $0.47^{+5.1\%}_{-4.7\%}\pm6.2\%$ & $0.53^{+2.1\%}_{-2.1\%}\pm2.1\%$ & $0.51^{+1.1\%}_{-0.6\%}\pm2.1\%$ \\
  \hline
  $500$ & $0.18^{+6.4\%}_{-5.8\%}\pm6.3\%$ & $0.19^{+2.1\%}_{-2.3\%}\pm2.3\%$ & $0.19^{+1.1\%}_{-0.8\%}\pm2.3\%$ \\
  \hline
  $600$ & $0.075^{+7.4\%}_{-6.6\%}\pm6.6\%$ & $0.080^{+2.2\%}_{-2.5\%}\pm2.5\%$ & $0.076^{+1.1\%}_{-0.9\%}\pm2.5\%$ \\
  \hline
\end{tabular}
\end{center}
\end{table}
\begin{table}
\caption{Same as Tab.\ \ref{tab:Selectron_table}, but for the pair production of maximally mixed staus.}
\label{tab:Stau_table_MM}
\begin{center}
\begin{tabular}{|c||c|c|c|}
  \hline
  $m_{\tilde{\tau}}$ [GeV] & LO (LO global) [fb] & NLO (NLO global) [fb] & NLO+NLL (id.\ global) [fb]\\
  \hline
  $100$ & $87.99^{+3.3\%}_{-4.2\%}\pm6.4\%$ & $112.83^{+2.1\%}_{-1.7\%}\pm1.6\%$ & $112.82^{+0.3\%}_{-0.9\%}\pm1.6\%$ \\
  \hline
  $200$ & $8.40^{+0.8\%}_{-1.1\%}\pm6.2\%$ & $9.99^{+2.2\%}_{-1.5\%}\pm1.7\%$ & $9.94^{+0.6\%}_{-0.4\%}\pm1.7\%$ \\
  \hline
  $300$ & $1.84^{+3.3\%}_{-3.3\%}\pm6.2\%$ & $2.10^{+2.1\%}_{-1.8\%}\pm1.9\%$ & $2.07^{+0.9\%}_{-0.4\%}\pm1.9\%$ \\
  \hline
  $400$ & $0.57^{+5.1\%}_{-4.7\%}\pm6.2\%$ & $0.63^{+2.1\%}_{-2.1\%}\pm2.1\%$ & $0.62^{+1.1\%}_{-0.6\%}\pm2.1\%$ \\
  \hline
  $500$ & $0.21^{+6.4\%}_{-5.8\%}\pm6.3\%$ & $0.23^{+2.1\%}_{-2.3\%}\pm2.3\%$ & $0.22^{+1.1\%}_{-0.8\%}\pm2.3\%$ \\
  \hline
  $600$ & $0.090^{+7.4\%}_{-6.6\%}\pm6.6\%$ & $0.096^{+2.2\%}_{-2.5\%}\pm2.5\%$ & $0.091^{+1.1\%}_{-0.9\%}\pm2.5\%$ \\
  \hline
\end{tabular}
\end{center}
\end{table}
The central NLO+NLL results have again been obtained with the $K$-factor
method, while the NLO+NLL (asymmetric) scale uncertainties have been
computed directly, and the PDF (symmetric) uncertainties at NLO using
the global NNPDF3.0 fits. The latter are therefore again identical in
the last two columns.

\section{Conclusion}
\label{sec:5}

To summarise, we have studied in this paper the effect of modern, NLO+NLL
PDFs on consistent NLO+NLL predictions for slepton pair production at Run
II of the LHC. Compared to previous work by us and other authors, we have
updated the analysis of left-handed selectron or smuon as well as
right-handed and maximally mixed stau pair production to the current LHC
centre-of-mass energy of 13 TeV. Also, cross sections at LO, NLO and
NLO+NLL have been computed with NNPDF3.0 PDFs from a global fit at NLO and
their uncertainties as estimated with the replica method, as well as from a fit
based on threshold-resummation improved NLO+NLL matrix elements of a reduced
set of observables (DIS, DY and top pair production). We applied a
factorisation method proposed previously that minimises the effect
of the data set reduction in the PDF fits, avoids the known problem of
outlier replicas, and preserves the reduction of the scale uncertainty
in our resummation calculation.

Apart from the generally known fact that hadronic cross sections
increase significantly at higher collision energies, we also observed
slightly larger cross sections, in particular for large slepton masses,
due to the NLO PDF update. We confirmed that the consistent use of
threshold-improved PDFs partially compensates resummation contributions
in the matrix elements. Together with the reduced scale uncertainty
at NLO+NLL, the described method further increases the reliability of
slepton pair production cross sections at the LHC. The new method has
been implemented for sleptons in the public code RESUMMINO.

\section*{Acknowledgements}
\noindent
We thank Marco Bonvini for useful discussions about the NNPDF3.0
resummation-improved PDFs. We also thank Valentina Dutta for providing
us with the parameter cards used for the CMS stau mixing scenarios
and Pieter Everaerts for the punctual clarifications about the CMS stau
analysis. This work has been supported by the BMBF under contract
05H15PMCCA and the DFG through the Research Training Network 2149
``Strong and weak interactions - from hadrons to dark matter''.

\bibliographystyle{apsrev4-1}
\bibliography{bib}

\end{document}